\documentclass[aps,prl,superscriptaddress,preprintnumbers,showpacs,twoside,twocolumn,floatfix]{revtex4}

\usepackage{bm}
\usepackage{graphicx}
\usepackage{amsmath}

\usepackage{amsmath, amssymb}

\begin{document}

\title{Thermometry of cold atoms in optical lattices via artificial gauge fields}

\author{Tommaso Roscilde}
\affiliation{Laboratoire de Physique, CNRS UMR 5672, Ecole Normale Sup\'erieure de Lyon, Universit\'e de Lyon, 46 All\'ee d'Italie, 
Lyon, F-69364, France}

\begin{abstract}
Artificial gauge fields are a unique way of manipulating the motional state of cold atoms. Here we propose the use (practical or conceptual) of artificial gauge fields -- obtained \emph{e.g.} via lattice shaking -- to perform primary noise thermometry of cold atoms in optical lattices - not requiring any form of prior calibration. The proposed thermometric scheme relies on fundamental fluctuation-dissipation relations, connecting the global response to the variation of the applied gauge field and the fluctuation of quantities related to the momentum distribution (such as the average kinetic energy or the average current). We demonstrate gauge-field thermometry for several physical situations, including free fermions and strongly interacting bosons. The proposed approach is extremely robust to quantum fluctuations - even in the vicinity of a quantum phase transition - when it relies on the thermal fluctuations of an emerging classical field, associated with the onset of Bose condensation or chiral order.    
\end{abstract}

\pacs{03.75.Lm, 42.50.Lc, 07.20.Dt, 03.75.Hh}

\maketitle

Cold atoms in optical lattices \cite{Blochetal2008, Lewenstein2012} currently represent the most prominent candidate for the quantum simulation \cite{qsimNature}  of lattice bosons and fermions, given the extreme level of tunability of the \emph{microscopic} Hamiltonian parameters (tunneling, interaction, etc.), and the rapidly developing toolbox of experimental probes to characterize the many-body quantum state \cite{Blochetal2008}. Yet one of the most important limitations of cold-atom quantum simulation is the intrinsic difficulty to control the \emph{macroscopic} parameters of the system, such as the chemical potential or the temperature, due to the fact that the system under investigation is virtually decoupled from any reservoir. This limitation is particularly serious if one is willing to reconstruct the equilibrium phase diagram of complex lattice Hamiltonians, study the nature of their phase transitions, etc. In particular, a proper quantum simulator should be equipped with a primary thermometric scheme, not needing any fit to theory data relative to the model of interest.  

Most prominent proposals for primary thermometry of strongly interacting cold atoms \cite{McKayDM2011} rely on the ability to image \emph{in situ} the atomic cloud \cite{ZhouH2011,Maetal2010}: regarding the (parabolic) trapping potential as a slowly varying external field coupling to the density, one can extract the local compressibility from the density gradient within a local-density approximation (LDA) scheme, and relate it to the local density fluctuations via a fundamental fluctuation-dissipation (FD) relation, which in turn enables to extract the temperature. This thermometry scheme (or a simplified version thereof) has been exploited in recent experiments equipped with a quantum-gas microscope \cite{Sanneretal2010,Muelleretal2010,Vogleretal2013}.
An alternative approach, also based on high-resolution imaging and LDA, implies a fit of the tails of the atomic cloud to the known thermodynamics of a diluted Bose (or Fermi) gas \cite{Shinetal2008,Shersonetal2010}. 
The requirement of high-resolution imaging can be often challenging in the experiments - especially when dealing with three-dimensional atomic clouds. Moreover the above thermometry scheme completely breaks down when using box traps - recently becoming available via holographic techniques \cite{Bakretal2009,  Shersonetal2010, Gauntetal2013} - which represent an important step towards the quantum simulation of bulk many-body phases. 
 
 In this Letter we propose a new thermometry scheme for optical-lattice quantum simulators, based on the use of artificial gauge fields (AGF), which have recently become available in the experiments \cite{Dalibardetal2010}. In particular a gauge field offers a unique tool to manipulate the momentum distribution, namely the most accessible observable in cold-atom experiments. Exploiting FD relations which link the variation of the momentum distribution upon applying a AGF to the noise in the momentum distribution itself, we devise a primary noise thermometer solely relying on time-of-flight measurements. In particular the gauge field necessary for thermometry can be trivial (namely it amounts to a simple gauge transformation of the Hamiltonian), in which case it does not even need to be realized experimentally, but it can be mimicked by a simple shift of the momentum distribution. More generally, the required gauge field for thermometry can be achieved via lattice shaking \cite{Lignieretal2007,Strucketal2011,Arimondoetal2012, Strucketal2012,Strucketal2013} - a scheme easily integrated in standard optical lattice experiments; an alternative scheme based on a combination of rf- and Raman fields has been realized in Ref.~\cite{Jimenez-Garciaetal2012}. 
 We demonstrate gauge-field thermometry for a variety of physical systems, showing its robustness to the presence of strong quantum fluctuations provided that one bases thermometry on the FD relation involving an order parameter, minimally affected by quantum fluctuations.
\begin{figure}
 \includegraphics[width=8cm]{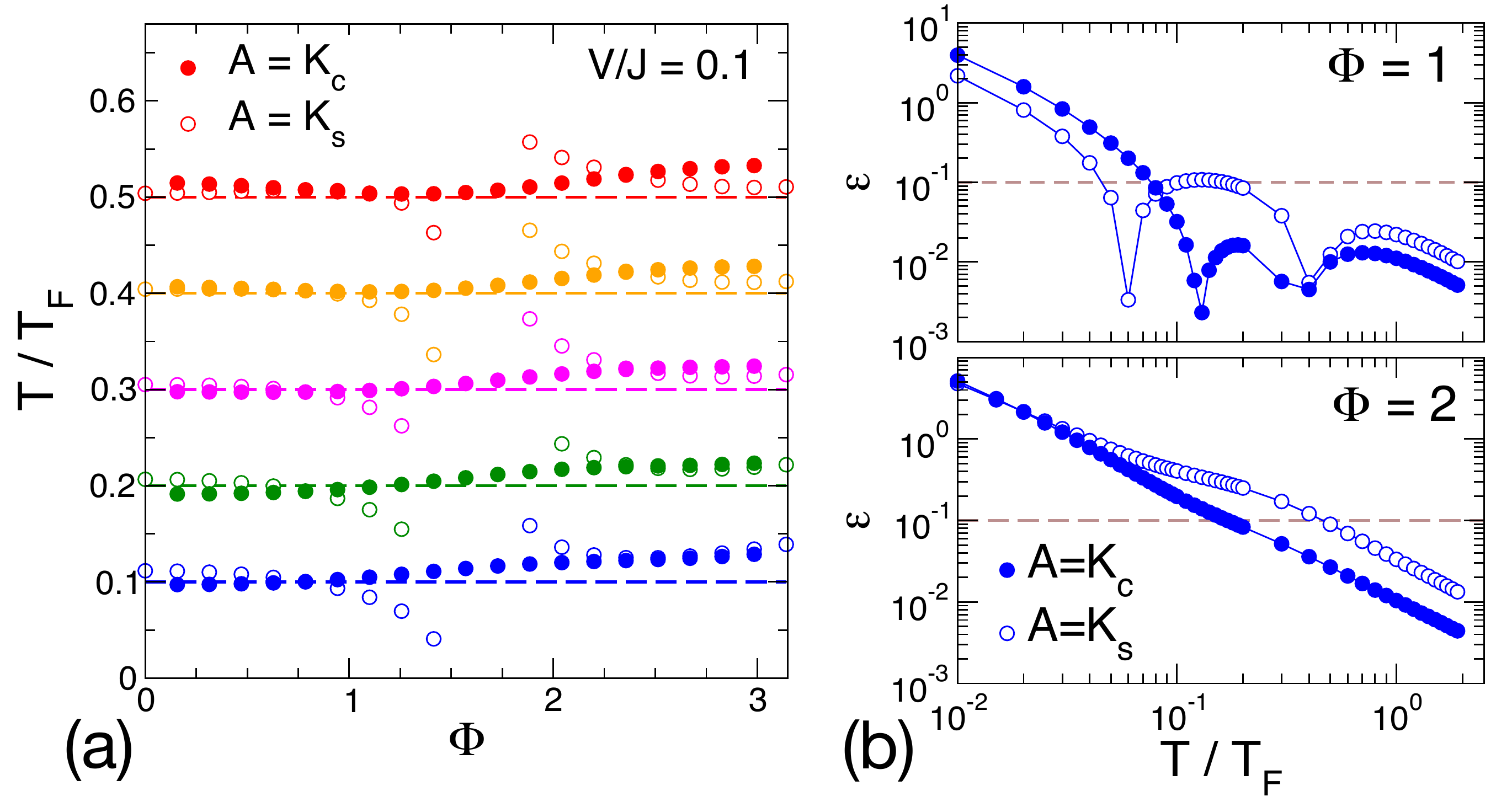}
 \caption{Gauge-field thermometry for free fermions on a triangular lattice. \emph{Left panel}. Estimated temperatures for $N=50$ fermions in a trapping potential $V_t/J = 0.1$ as a function of the staggered flux $\Phi$.       
\emph{Right panel.} Relative error of the temperature estimator, $\epsilon = (T_A-T)/T$, as function of temperature and for two different $\Phi$ values -- other parameters as in the previous panel. The horizontal dashed line marks the $10\%$ accuracy threshold.}
\label{f.fermions}
 \end{figure} 
   \begin{figure*}[ht!]
 \includegraphics[width=16.5cm]{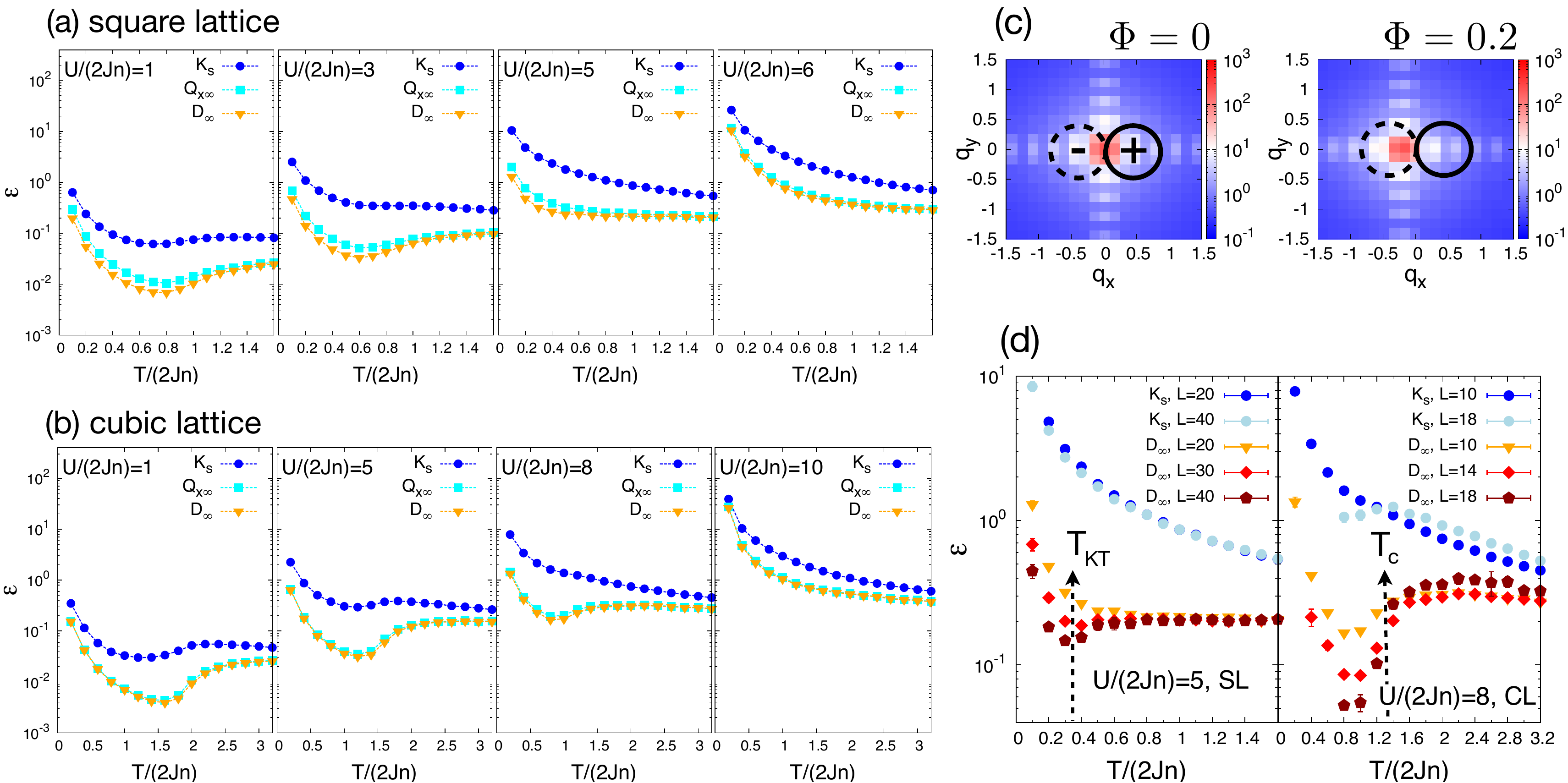}
 \caption{Gauge-field thermometry for the quantum rotor (QR) model on the square lattice (SL) and cubic lattice (CL). (a)-(b) Thermometry accuracy related to the observables $K_s$, $Q_{x\infty}$ and $D_{\infty}$ (see text) as a function of temperature for a SL with $L=20$ (a) and a CL with $L=10$. (c) $\Phi$-dependence of the peak position of a QR model on the SL ($L=20$). The solid and dashed circles refer to the two contributions to $D_{\infty}$. (d) Scaling of thermometry accuracy for the QR model close to the Mott insulator quantum critical point; the arrows mark the critical temperatures $T_{\rm KT}/(2Jn) = 0.34(2)$ (SL) and $T_{\rm c}/(2Jn) = 1.35(5)$ (SL).}
\label{f.squarecubic}
 \end{figure*}

 For definiteness, let us consider the following general Hamiltonian for quantum particles in an artificial gauge field ${\cal H} = {\cal H}_J + {\cal H}_U + {\cal H}_t$. Here
 \begin{equation}
 {\cal H}_J = -J \sum_{\langle ij \rangle~ || ~x } \left( e^{i\Phi} a_i^{\dagger} a_j + {\rm h.c.} \right) -J \sum_{\langle lm \rangle ~||~ \bar{x}} \left(a_l^{\dagger} a_m + {\rm h.c.} \right) 
 \label{e.HJ}
 \end{equation}  
 describes the hopping between pairs of nearest-neighboring sites $\langle ij \rangle$ and $\langle lm \rangle$. In particular we assume the presence of a uniform Peierls phase $\Phi$ on all the bonds 
 $\langle ij \rangle$ parallel to a given lattice direction ($x$), while no gauge field is present along the other directions ($\bar{x}$). Such a gauge field does not create a finite magnetic flux in a square, hexagonal or cubic lattice, but it creates a staggered flux in a triangular lattice (as recently demonstrated in Ref.~\onlinecite{Strucketal2013}, although in a different gauge), in a Kagom\'e lattice etc. 
 Furthermore ${\cal H}_U$ represents an on-site interaction term, and ${\cal H}_t$ a trapping potential term. In the following we will consider both the case of fermionic as well as bosonic operators $a$, $a^{\dagger}$.

  In our scheme the Peierls phase $\Phi$ plays the role of the probe field. The response of a generic observable $\langle A \rangle$ ($\langle... \rangle$ denoting the statistical average) to a variation of the applied GF is given by
  \begin{equation}
  \frac{\partial \langle A\rangle} {\partial \Phi} = \frac{J}{T} \left[\cos\Phi~ {\rm cov}_{\tau} (A,K_s) - \sin\Phi~ {\rm cov}_\tau(A,K_c)\right]~.
  \label{e.FD}
  \end{equation} 
 Here
 \begin{eqnarray} 
 K_c = \sum_{\langle ij \rangle || x } \left(a_i^{\dagger} a_j + {\rm h.c.} \right) ~~~~~~
 K_s = \i \sum_{\langle ij \rangle || x } \left(a_i^{\dagger} a_j - {\rm h.c.} \right)
 \end{eqnarray}
 are respectively the Josephson coupling and the current operator along the $x$ direction, and ${\rm cov}_{\tau}$ is the time-averaged covariance,
 ${\rm cov}_{\tau} (A,B) = \frac{1}{\beta} \int_0^{\beta} d\tau \left[ \langle A(0) B(\tau) \rangle - \langle A \rangle \langle B \rangle\right]$ ($\beta=T^{-1}$). 
 Throughout the paper we set $k_B=1$.
 
The capability of applying a tunable GF $\Phi$ \cite{Strucketal2011, Strucketal2013, Jimenez-Garciaetal2012} allows to reconstruct experimentally the response function $\partial \langle A\rangle / \partial \Phi$ to the l.h.s of Eq.~\ref{e.FD}. Thermometry can be achieved through Eq.~\ref{e.FD} if one is also able to measure the r.h.s., containing the correlation between the fluctuations of the observable A and of the operators $K_c$, $K_s$ -- which can be extracted experimentally as discussed below. It is obvious that experiments cannot reconstruct the time-averaged covariance, but rather the equal-time one, ${\rm cov}(A,B) = \langle A B \rangle - \langle A \rangle \langle B \rangle$. If $A$ (or $B$) commutes with the Hamiltonian, then ${\rm cov}_{\tau}(A,B) = {\rm cov}(A,B)$; yet for a generic optical lattice experiment the only integrals of motion are the total particle number (which is independent of the applied flux) and the total energy (which is not easily measurable). Hence one has to resort to a judiciously chosen observable $A$, whose quantum fluctuations are well controlled - we will discuss in the following how to choose such an observable. For any observable $A$, one has a \emph{temperature estimator} $T_A$ which is accessible experimentally
\begin{equation}
\frac{T_A}{J} = : \frac{\cos\Phi~ {\rm cov} (A,K_s) - \sin\Phi~ {\rm cov} (A,K_c)} {\partial_{\Phi} \langle A \rangle}~.
\label{e.TA}
\end{equation}
Beyond its practical thermometry scope, the temperature estimator $T_A$ has also a fundamental meaning: it expresses an effective temperature related to the observable $A$ in question, accounting for its combined thermal \emph{and} quantum fluctuations. We will elaborate further on this point towards the end of the paper. 

The operators $K_c$ and $K_s$ can be easily extracted from the momentum distribution operator 
$n({\bm k})  = |w({\bm k})|^2 \frac{1}{V} \sum_{ij} e^{i{\bm k}({\bm r}_i - {\bm r}_j)}  a_i^{\dagger} a_j $ via an inverse Fourier transform (here $w({\bm q})$ is the Fourier transform of the lattice Wannier function):
\begin{eqnarray}
K_c &=& \frac{V}{(2\pi)^d} \int d^d q~ \cos({\bm q}\cdot \hat{\bm x})~ n^{(+)}(\bm q) \nonumber \\
K_s &=& \frac{V}{(2\pi)^d} \int d^d q~ \sin({\bm q}\cdot \hat{\bm x})~ n^{(-)}(\bm q)
\label{e.KsKc}
\end{eqnarray}
where $n^{(\pm)}({\bm q}) = \left[n({\bm q}) \pm n(-q_x,q_y,q_z)\right]/2$ is the momentum distribution symmetrized/antisymmetrized with respect to a reflection about the $yz$ plane. Therefore each time-of-flight measurement realizes a \emph{projective measurement} of $K_c$ and $K_s$; if the observable $A$ is also a quantity measured projectively via time-of-flight, a generic cold-atom experiment can access the full-counting statistics of all these quantities \cite{footnotePSF}, and in particular the covariances contained in the temperature estimator, Eq.~\eqref{e.TA}. Moreover $A$ should be such that $\partial_{\Phi}\langle A \rangle \neq 0$, and in particular it is convenient that the $\Phi$-dependence of $\langle A \rangle$ be significant around the value of $\Phi$ of interest. 
 
 \emph{Free fermions on the triangular lattice.} We begin by illustrating the thermometry scheme in question in the case of free fermions (${\cal H}_U=0$). In the absence of a trapping potential the Hamiltonian commutes with $K_c$ and $K_s$, so that $T_{K_c} = T_{K_s} = T$. Yet the presence of a trapping potential term, ${\cal H}_t = V \sum_i ({\bm r}_i-{\bm r}_0)^2 n_i$ induces quantum fluctuations in the $K_s$ and $K_c$ operators, giving rise to a discrepancy between the above temperature estimators and the actual temperature. To evaluate the actual impact of quantum fluctuations, we consider a diluted fermionic system of $N=50$ fermions on a triangular lattice, subject to a strong confining potential $V = 0.1J$. Fig.~\ref{f.fermions} shows exact diagonalization results (see \cite{suppmat}) for the temperature estimators: we observe that for most values of the flux $\Phi$ both $T_{K_c}$ and  $T_{K_s}$ lie very close to the actual temperature, and that the relative error $\epsilon = |T_A - T|/T$ is typically $<10\%$ for $T> 0.1 T_F$; we anticipate a significantly better accuracy of the temperature estimators for systems which experience a weaker confining potential. It is important to observe that $T_{K_s}$  has a sharp singularity at 
 $\Phi = \pi/2$, due to the fact that $\partial_\Phi \langle K_s \rangle$ vanishes at that value \cite{caution}. On the other hand,  $\partial_\Phi \langle K_c \rangle = 0$ for $\Phi= 0, \pi$. Therefore the two temperature estimators are fully complementary \cite{footnotederiv}. 
 
 \emph{Strongly interacting bosons.} For the rest of the paper we focus our attention on strongly interacting bosons undergoing a transition from a superfluid (SF) phase to a Mott insulator (MI) phase.  
The numerical study of the Hamiltonian with frustrated kinetic energy Eq.~\eqref{e.HJ} is notoriously difficult, given that the complex hopping amplitudes produce a negative sign problem in quantum Monte Carlo simulations. Nonetheless in the limit of large integer filling $n \gg 1$ the Bose-Hubbard model admits a mapping onto the quantum rotor Hamiltonian \cite{Wallinetal1994, Teichmannetal2009}:
\begin{eqnarray}
{\cal H}_{\rm QR} &=& -2Jn \sum_{\langle ij \rangle~ || ~x } \cos(\phi_i-\phi_j-\Phi)  \nonumber \\
&&- 2Jn \sum_{\langle lm \rangle ~||~ \bar{x}} \cos(\phi_l-\phi_m) - \frac{U}{2} \sum_i \frac{\partial^2}{\partial \phi_i^2}~.
\label{e.QR}
 \end{eqnarray} 
 where $\phi_i$ is the local phase of the lattice Bose operator. 
Given the constraint on integer filling, the effect of trapping cannot be faithfully reproduced within the quantum rotor Hamiltonian - yet we mimic the effect of confinement by introducing open boundary conditions. The quantum rotor Hamiltonian conveniently lends itself to path-integral Monte Carlo (PIMC) simulations \cite{Wallinetal1994}, which, being formulated in the basis of the phase eigenstates, allow to reconstruct the full-counting statistics of any quantity related to the momentum distribution, much as in the time-of-flight experiments. Moreover the quantum rotor formulation allows for the introduction of arbitrary gauge fields without the appearance of a sign problem. Simulations are performed on $L\times L$ square lattices and $L^3$ cubic lattices - further details can be found in \cite{suppmat}.

\emph{Square and cubic lattice.} We begin our discussion with the case of square and cubic lattices, for which the Peierls phase introduced in Eq.~\ref{e.HJ} are gauge-trivial. Yet they represent an invaluable \emph{conceptual} thermometry tool in the superfluid regime at \emph{zero} gauge field for both lattices. 
The tool is purely conceptual because the Peierls phase induces a simple shift of the momentum distribution $n({\bm q}) \to n({\bm q} + \Phi \hat{x})$ (Fig.~\ref{f.squarecubic}(c)) so that experimentally the actual application of the Peierls phase is \emph{not} needed -- all $\Phi$-derivatives can be evaluated by simply translating the measured momentum distribution. 
We consider the response of several $n({\bm q})$-related quantities to the applied Peierls phase: 1) the already mentioned current operator $K_s$; 
2) a measure of the asymmetry of the momentum distribution peak at $q=0$, 
\begin{equation}
D_{\infty} = (N_+ - N_-)/(N_+ + N_-);
\end{equation} 
here  $N_{\pm} = \sum_{{\bm q} \in {\cal D}_{\pm}} n({\bm q})$ is the number of particles in a disk ${\cal D}_{\pm}$  in momentum space of radius $R = 0.4 a^{-1}$ (where $a$ is the lattice spacing), centered around the point ${\bm q}_{\pm} = (\pm R, 0)$ (see Fig.~\ref{f.squarecubic}(c) for an illustration) \cite{footnoteD3d};
3) the total $x$-component of momentum in the double-disk region, $Q_x,\ = \sum_{{\bm q} \in {\cal D}_{\pm}} q_ x n({\bm q})/(N_+ + N_-)$, namely the first moment (along the $x$-direction) of the momentum distribution on the two-disk region.  
The three quantities are chosen so as to have a non-vanishing derivative as a function of $\Phi$. In particular $D_{\infty}$ and $Q_{x\infty}$ are sensitive to the displacement of the $q=0$ peak upon changing the gauge field \cite{footnotepeak}. 

Fig.~\ref{f.squarecubic}(a-b) shows the temperature dependence of the thermometry relative error $\epsilon$ based on Eq.~\eqref{e.TA} with  $A=K_s$, $D_{\infty}$ and $Q_{x\infty}$ for different values of the boson-boson interaction $U$, up to and past the critical point for superfluid/Mott-insulator (SF/MI) transition - which is estimated \cite{Teichmannetal2009} for quantum rotors on the square lattice to be $[U/(2Jn)]_c \approx 5.8$, and on the cubic lattice to be  $[U/(2Jn)]_c \approx 10$ . We observe that the relative error degrades as one increases the interaction, due to the enhanced quantum fluctuations in the momentum distribution. Yet the temperature dependence of $\epsilon$ is highly non-monotonic, showing a distinct dip at intermediate temperatures, particularly marked for thermometry based on the two quantities related to the condensate peak, namely $D_{\infty}$ and $Q_{x\infty}$. At first sight this might appear surprising, as the low-temperature regime is the one in which quantum fluctuations jeopardizing the accuracy of the proposed thermometry are most prominent. On the other hand, it is immediate to relate this anomaly in the temperature behavior of $\epsilon$ to the phase transition of the two systems under investigation, namely the Kosterlitz-Thouless (KT) transition of the square lattice and the condensation transition on the cubic lattice. At and below the transition, the condensate fraction becomes macroscopic (in $d=3$) or quasi-macroscopic (in $d=2$), diverging in the thermodynamic limit, while its fluctuations (both thermal \emph{and} quantum) do not follow the same scaling. This means that thermometry related to fluctuations of the condensate peak can be expected to be minimally affected by quantum fluctuations. In particular, as shown in Fig.~\ref{f.squarecubic}(d), upon increasing the size of the system, the accuracy of thermometry \emph{improves} with growing system size below the KT transition in 2$d$ ($T<T_{\rm KT}$) and below the condensation transition in 3$d$ ($T<T_c$).
A detailed scaling analysis \cite{suppmat} shows that the relative error $\epsilon$ of thermometry associated with $D_{\infty}$ appears to scale as $L^{-1}$ below the KT transition in 2$d$ and as $(L \log L)^{-1}$ below the condensation transition in 3$d$, as a result of the scaling of the covariance difference 
$\Delta{\rm cov} (D_{\infty},K_s) = {\rm cov}(D_{\infty},K_s) - {\rm cov}_{\tau}(D_{\infty},K_s)$ as well as of the derivative $\partial_{\Phi} D_{\infty}$ (the latter showing agreement with Bogolyubov theory). On the other hand, the relative error ceases to scale in the normal phase. 

 \begin{figure}[h!]
 \includegraphics[width=9cm]{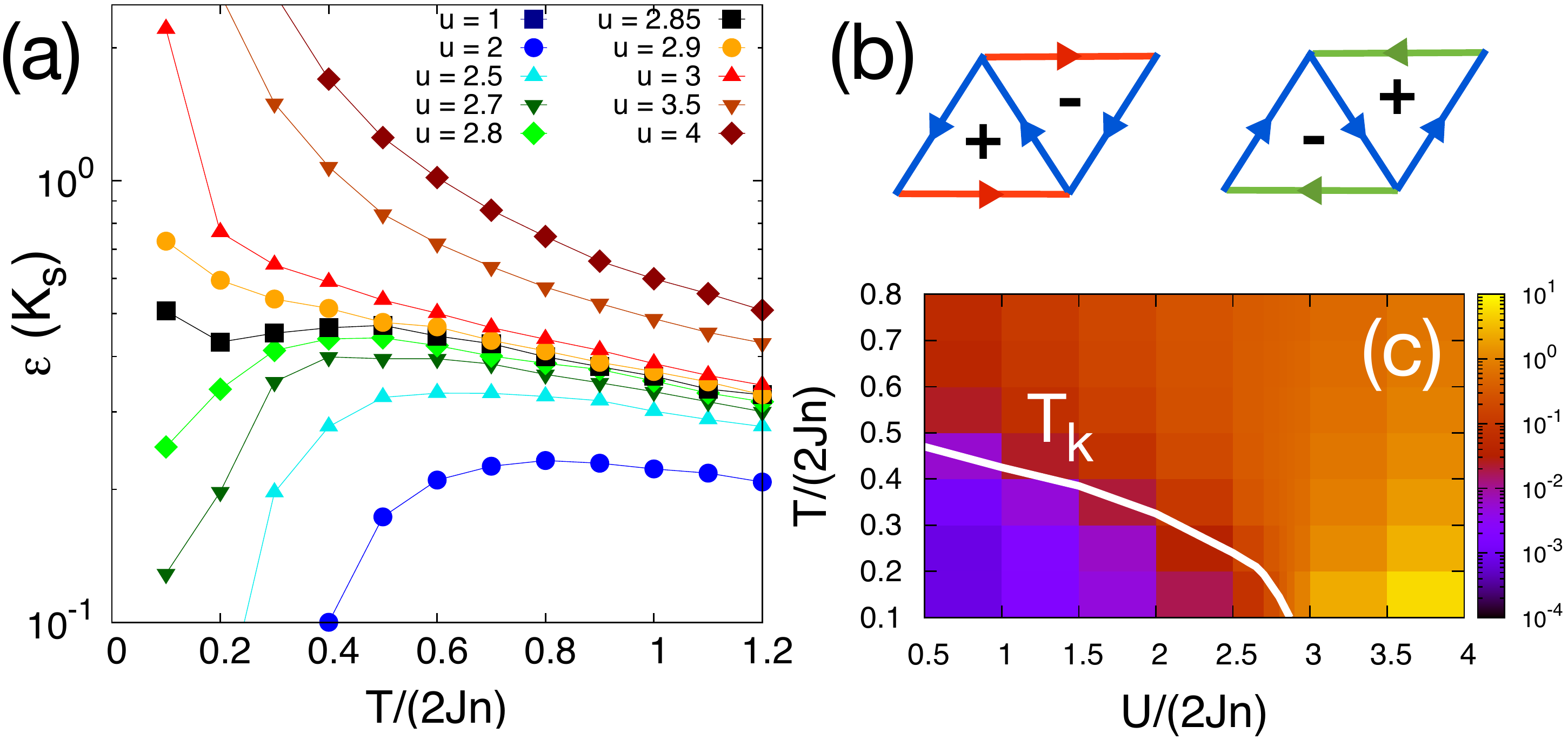}
 \caption{Gauge-field thermometry for the QR model on a triangular lattice with a $\pi$-flux. (a) $T$-dependence of the accuracy of thermometry based on the current $K_s$ for different interactions $u = U/(2Jn)$ crossing the quantum critical point $u_c \approx 2.8$. (b) (c) Phase diagram of the model, with comparison between the chiral transition temperature $T_k$ and the $\epsilon(K_s)$ accuracy (false colors).}
\label{f.triangular}
 \end{figure}   
 
 \emph{Triangular lattice with a $\pi$-flux.} As a last example, we consider thermometry in a system with an already applied gauge field, namely a triangular lattice with a $\pi$-flux, realized recently via lattice shaking in Refs.~\cite{Strucketal2011, Strucketal2013}.  Such a system possesses an additional, discrete $Z_2$ symmetry associated with the choice of two degenerate vortex patterns (with alternation of vortices/antivortices on adjacent plaquettes - Fig.~\ref{f.triangular}(b) \cite{Leeetal1984}). This symmetry is broken via a chiral Ising transition at a critical temperature $T_k$, as also observed in the experiments \cite{Strucketal2011, Strucketal2013}. In particular the two vortex patterns can be distinguished by the appearance of a finite current $K_s$ on the horizontal links of the triangular lattice, serving therefore as an order parameter. Fig.~\ref{f.triangular}(a,c) shows the accuracy of thermometry based on $K_s$ as a function of temperature and of increasing repulsion $U$ among the bosons. For $U>U_c \approx 2.8 *(2Jn)$ the repulsion drives the system across a quantum critical point QCP with destruction of long-range chiral order \cite{RoscildeTBP}. Yet we observe that, similarly to the condensation transition of the previous example, the chiral transition boosts the accuracy of thermometry in a dramatic fashion up to the QCP. 
 In particular, in the chiral phase the relative error of the proposed thermometry scales as $L^{-2}$, driven by the very violent divergence of $\partial_{\Phi} \langle K_s \rangle$ (as $L^4$)\cite{suppmat}.
 
In conclusion, we propose gauge-field thermometry for cold atoms in optical lattice, namely primary thermometry based on monitoring the fluctuations of the momentum distribution, as well as its response to the application of an artificial gauge field. In the case of strongly interacting bosons, it is shown to provide absolute thermometry in the thermodynamic limit when the system develops condensation or of chiral order.  
Our method does not require any form of microscopy, nor the presence of a smooth and well-characterized confining potential. 
 The proposed thermometry paves the way for the reconstruction of the phase diagram of fundamental lattice models using cold-atom quantum simulators. Future work \cite{RoscildeTBP} will address the extension of this scheme to continuum space as well as to disordered systems. 
 

 Stimulating discussions with J. Simonet, F. Gerbier and especially A. Eckardt are gratefully acknowledged, as well as support from the PSMN (ENS Lyon), where all calculations were performed.

\newpage

\setcounter{page}{1}
\setcounter{figure}{0}
\setcounter{equation}{0}

\section{Supplementary Material for "Thermometry of cold atoms in optical lattices via artificial gauge fields"}

Here we provide: 1) a discussion of the effect of finite momentum resolution in the time-of-flight images; 2) the technical details of the numerical simulations leading to the results presented in the paper; 3) a scaling analysis of the gauge-field thermometry for the square, cubic and $\pi-$flux triangular lattice. 

\subsection{Effect of finite momentum resolution in time-of-flight images}

In the following we assume that the accuracy in the experimental measurement of the momentum distribution is dominated by the limitations of the imaging system. Other sources of error are the finite time of flight (as the momentum distribution is only recovered in the far-field limit) and the interaction effects during time of flight. While the latter can be in principle removed via the use of Feshbach resonance, the former is found theoretically to have modest effects compared to imaging resolution \cite{Gerbieretal2008}. 

We assume that the experimentally measured momentum distribution integrated along the line of sight, $\bar{n}^{\rm (exp)}({\bm q})$ is linearly related to the actual one $\bar{n}({\bm q})$ as 
\begin{equation}
\bar{n}^{\rm (exp)}({\bm q}) = \int d^2 q' f_{\rm PSF}(|{\bm q}-{\bm q}'|) ~\bar{n}({\bm q}')
\end{equation}
where $f_{\rm PSF}(|{\bm q}-{\bm q}'|)$ is the point-spread function (PSF) associated with the imaging resolution. 

 Making using the convolution theorem, the experimentally measurable values of the Josephson and current operators, Eqs.~\eqref{e.KsKc} are linearly related to their actual values via
 \begin{equation}
 K_{c,s}^{\rm (exp)} = G_f~ K_{c,s}
\end{equation}
where 
\begin{equation}
G_f  = \int d^2 q ~f_{\rm PSF}(|\bm q|) ~e^{-i{\bm q}\cdot{\bm a_1}} 
\end{equation}
is the inverse Fourier transform of the point-spread function (PSF). 

As a consequence the experimentally measured temperature estimators, Eq.~\eqref{e.TA},  read
\begin{equation}
T_{A}^{\rm (exp)} = G_f~ T_{A}~.
\end{equation}
If the PSF is  well characterized, the $G_f$ factor can be corrected for in the temperature estimator.  

In the specific case of the temperature estimators related to $D_{\infty}$ and $Q_{x\infty}$, the observables in question need to be redefined to match with the actually measured ones, namely
\begin{equation}
D_{\infty}^{\rm (exp)} = \frac{N^{(\rm exp)}_+ - N^{(\rm exp)}_-}{N^{(\rm exp)}_+ + N^{(\rm exp)}_-}  
\end{equation}
and
\begin{equation}
 Q^{(\rm exp)}_{x\infty} = \frac{\sum_{\bm q \in {\cal D}_{\pm}} q_x n^{\rm (exp)}({\bm q})}{N^{(\rm exp)}_+ + N^{(\rm exp)}_-}~ 
\end{equation} 
where we have defined
\begin{equation}
 N^{(\rm exp)}_{\pm} = \sum_{\bm q \in {\cal D}_{\pm}} n^{\rm (exp)}({\bm q})~. 
\end{equation}

In the case of interacting bosons, it is natural to expect that thermometry based upon $D^{\rm (exp)}_{\infty}$ and $Q^{(\rm exp)}_{x\infty}$ has the same accuracy (and in particular the same scaling of the relative error with system size, see later) as the theoretical quantities defined in the main text. 

\subsection{Gauge-field thermometry for free fermions}
In this section we provide the main formulas for the calculation of the temperature estimators for free fermions. The solution of the single-particle problem in a trap, ${\cal H}_{J} + {\cal H}_t$, produces eigenvalues $\epsilon_{\alpha}$ and corresponding lattice eigenfunctions $\phi_{\alpha}(i)$. We make use of the 2-point and 4-point correlation functions for free fermions:
\begin{equation}
\langle a_i^{\dagger} a_j \rangle = \sum_\alpha \phi^*_{\alpha}(i) \phi_{\alpha}(j) ~f_T(\epsilon_\alpha)
\end{equation}
and 
\begin{equation}
\langle a_i^{\dagger} a_j  a_k^{\dagger} a_l \rangle = \sum_{\alpha\beta\gamma\eta} \phi_\alpha^*(i) \phi_\beta(j) \phi_\gamma^*(k) \phi_\delta(l) ~F_{\alpha\beta\gamma\eta}(ijkl)
\end{equation}
with 
\begin{eqnarray}
F_{\alpha\beta\gamma\delta}(ijkl) & = & \delta_{\alpha\beta} \delta_{\gamma\eta} (1-\delta_{\alpha\gamma}) f_T(\epsilon_\alpha) f_T(\epsilon_\gamma) \nonumber \\
&+& \delta_{\alpha\eta} \delta_{\beta\gamma} (1-\delta_{\alpha\eta}) f_T(\epsilon_\alpha)[1- f_T(\epsilon_\beta)] \nonumber \\
&+& \delta_{\alpha\beta} \delta_{\gamma\eta} \delta_{\beta\gamma} f_T(\epsilon_\alpha)
\end{eqnarray}
where $f_T(\epsilon_{\alpha}) = \left[e^{(\epsilon_{\alpha}-\mu)/(T)}+1\right]^{-1}$ is the Fermi distribution, and $\mu$ is the chemical potential. The latter expressions readily allow to calculate the expectation values of the Josephson and current operators
\begin{equation}
\langle K_{c,s} \rangle = \sum_{\alpha} K_{c,s}^{(\alpha\alpha)} f_T(\epsilon_\alpha) 
\end{equation}
where
\begin{eqnarray}
K_c^{\alpha\beta} &= & \sum_{\langle ij \rangle} \left[\phi^*_{\alpha}(i) \phi_{\beta}(j) + {\rm c.c.} \right]  \nonumber \\
K_s^{\alpha\beta} &= & \sum_{\langle ij \rangle} \left[\i \phi^*_{\alpha}(i) \phi_{\beta}(j) + {\rm c.c.} \right];
\end{eqnarray}
as well as the averages of $K_c^2$, $K_s^2$ and $K_s K_c$, obtained via the general formula
\begin{eqnarray}
\langle K_r K_{r'} \rangle &=& \sum_{\alpha\beta} K_r^{(\alpha)}  K_{r'}^{(\beta)} \nonumber \\
&+& \sum_{\alpha\beta} K_{r}^{(\alpha\beta)} K_{r'}^{(\beta\alpha)}
f_T(\epsilon_\alpha) \left[1- f_T(\epsilon_\beta)\right] 
\end{eqnarray}
where $r, r' = c, s$. This provides all the ingredients to build the variances and covariances ${\rm var}(K_s)$, ${\rm var}(K_c)$, and ${\rm cov}(K_c,K_s)$; the derivatives $\partial_{\Phi} \langle K_{c,s} \rangle$ are calculated numerically. This allows to reconstruct the temperature estimators $T_A$ (for $A = K_c, K_s$) as in Eq.~\eqref{e.TA} of the main text.

\subsection{Path-integral Monte Carlo simulations of the quantum rotor model}

The path-integral Monte Carlo (PIMC) simulations of the quantum rotor model presented in the main text follow the formulation of Ref.~\cite{sWallinetal1994}, and are based on a Trotter-Suzuki decomposition of the density operator with imaginary-time step $\Delta\tau = (MT)^{-1}$ such that
$\Delta\tau U = 10^{-2}$; under this condition the Trotter error is found to be essentially masked by the statistical error of the Monte Carlo data. In the simulation presented in the text, this amounts to Trotter dimensions with $M \sim 10^3$ steps for the lowest temperatures and strongest interactions considered. The lattice sizes we used in real space are instead $L=10,..., 60$ in $d=2$ and $L=6,..., 18$ in $d=3$.

The Trotter-Suzuki decomposition maps the quantum rotor model, Eq.~\eqref{e.QR} of the main text, onto a $(d+1)-$dimensional XY model with effective Hamiltonian
\begin{eqnarray}
\frac{{\cal H}_{\rm eff}}{2Jn} &=& -\frac{1}{M} \sum_{\langle ij \rangle~ || ~x,k } \cos(\phi_{i,k}-\phi_{j,k}-\Phi)  \nonumber \\
&&- \frac{1}{M} \sum_{\langle lm \rangle ~||~ \bar{x}} \cos(\phi_{l,k}-\phi_{m,k}) \nonumber \\
&& -  \frac{T}{\epsilon U} \sum_{i,k} \cos(\phi_{i,k}-\phi_{i,k+1}) 
\label{e.Heff}
 \end{eqnarray} 
where $(i,k)$ are the coordinates of a lattice site in the $(d+1)-$dimensional lattice. 

The statistical averages related to the effective Hamiltonian, Eq.~\eqref{e.Heff}, are obtained via a Monte Carlo algorithm based on a composite update scheme, consisting of: 1) a Metropolis single-spin rotation; 2) an over-relaxation move; 3) a Metropolis \emph{single-chain} update, proposing a rotation of the chain of spins having all the same spatial coordinate $i$; 4) a Wolff cluster algorithm. In particular the move 3) turns out to be very effective, given that the coupling in the imaginary-time dimension  $\frac{T}{\epsilon U}$ can be several orders of magnitude larger than the one in the real-space dimensions, inhibiting the individual rotation of single spins. In the case of the frustrated $\pi-$flux triangular lattice, the Wolff algorithm generates very large clusters percolating across the lattice even in the disordered phase, as generally expected in frustrated systems \cite{LeungH1991}. Yet the effective Hamiltonian, Eq.~\eqref{e.Heff}, has frustrated couplings only along $d$ dimensions while the $(d+1)-$th dimension is not frustrated, and therefore the cluster algorithm can be sensitive to the actual correlations in imaginary time. In general we find the geometry of the clusters to be non-trivial -- they are far from occupying the total volume of the lattice -- and, as a consequence, the cluster move is found to accelerate the dynamics in an appreciable manner.  

 The critical temperatures for quasi(-condensation) have been estimated from the scaling of the $n(\bm q=0)$ peak: $n(\bm q=0)\sim L^{7/4}$ at the KT transition, and 
 $n(\bm q=0)\sim L^{\gamma/\nu}$ with $\gamma/\nu\approx 1.98$ \cite{PelissettoV2002} at the condensation transition. The critical temperature for the chiral transition has been estimated via the divergence of the chirality structure factor, $S_{\kappa} = \frac{1}{L^2} \sum_{\triangle \triangle'} \langle \kappa_\triangle \kappa_\triangle' \rangle$: here the sum runs over up-pointing triangular plaquettes of the triangular lattice, and 
 \begin{eqnarray} 
\kappa_\triangle & = & \i (b^{\dagger}_1 b_2 + b^{\dagger}_2 b_3 +   b^{\dagger}_3 b_1) + \rm{h. c.} \nonumber \\ 
& \approx & n \left[ \sin(\phi_1-\phi_2) + \sin(\phi_2-\phi_3) + \sin(\phi_3-\phi_1)\right ]  \nonumber
 \end{eqnarray}
 where the indices $1$, $2$ and 3 denote the sites of a triangular plaquette (oriented counterclockwise), and the last expression of the above equation holds in the quantum rotor approximation.

\subsection{Scaling of thermometry accuracy for strongly interacting bosons}
\label{s.scaling}
 
In this section we analyze the scaling of the relative error of gauge-field thermometry in the case of strongly interacting bosons in the quantum rotor regime (large integer filling). For the case $\Phi=0, \pi$ of interest here, the relative error associated with the temperature estimator $T_A$ of Eq.~\eqref{e.TA} of the main text reads
\begin{equation}
\epsilon(A) = \frac{|T_A - T|}{T} = \frac{|\Delta {\rm cov} (A,K_s)|}{T~|\partial_\Phi\langle A \rangle|}~.  
\label{e.epsilon}
\end{equation}
Here $\Delta {\rm cov} = {\rm cov} - {\rm cov}_{\tau}$. is the difference between equal-time and time-averaged covariances.
In order to investigate the scale dependence of $\epsilon(A)$ we try and extract analytically and/or numerically the scaling of the numerator and denominator of Eq.~\eqref{e.epsilon}. 

\begin{figure}[h!]
 \includegraphics[width=8cm]{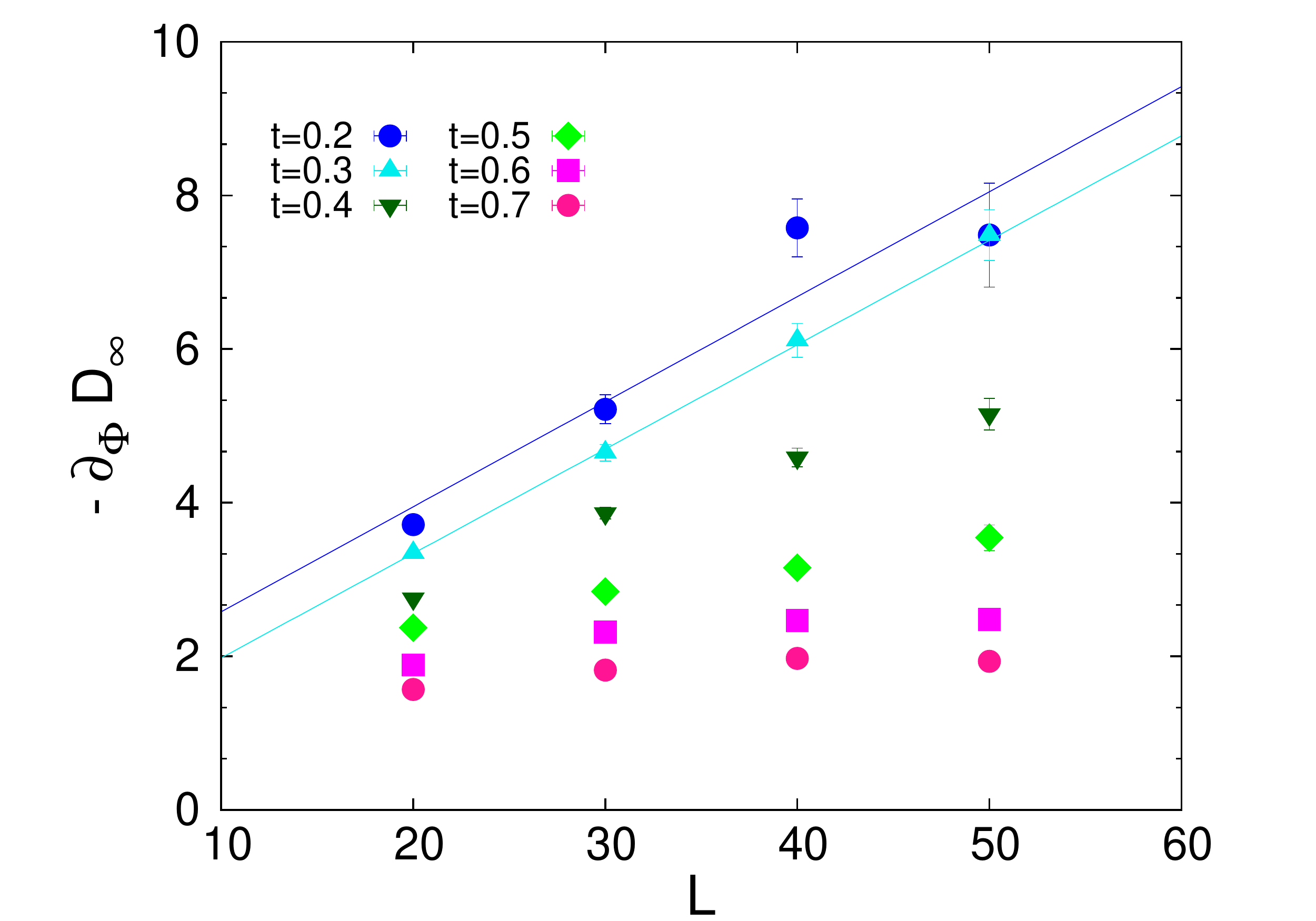}
 \caption{Scaling of $\partial_{\Phi} \langle D_{\infty} \rangle$ for the quantum rotor model on the square lattice with $U/(2Jn) = 5$. Here $T_{\rm KT}/(2Jn) =0.34(2)$. Solid lines are fits to the form $aL+b$.}
\label{f.Dnorm-SLscaling}
 \end{figure}   

\begin{figure}[h!]
 \includegraphics[width=8cm]{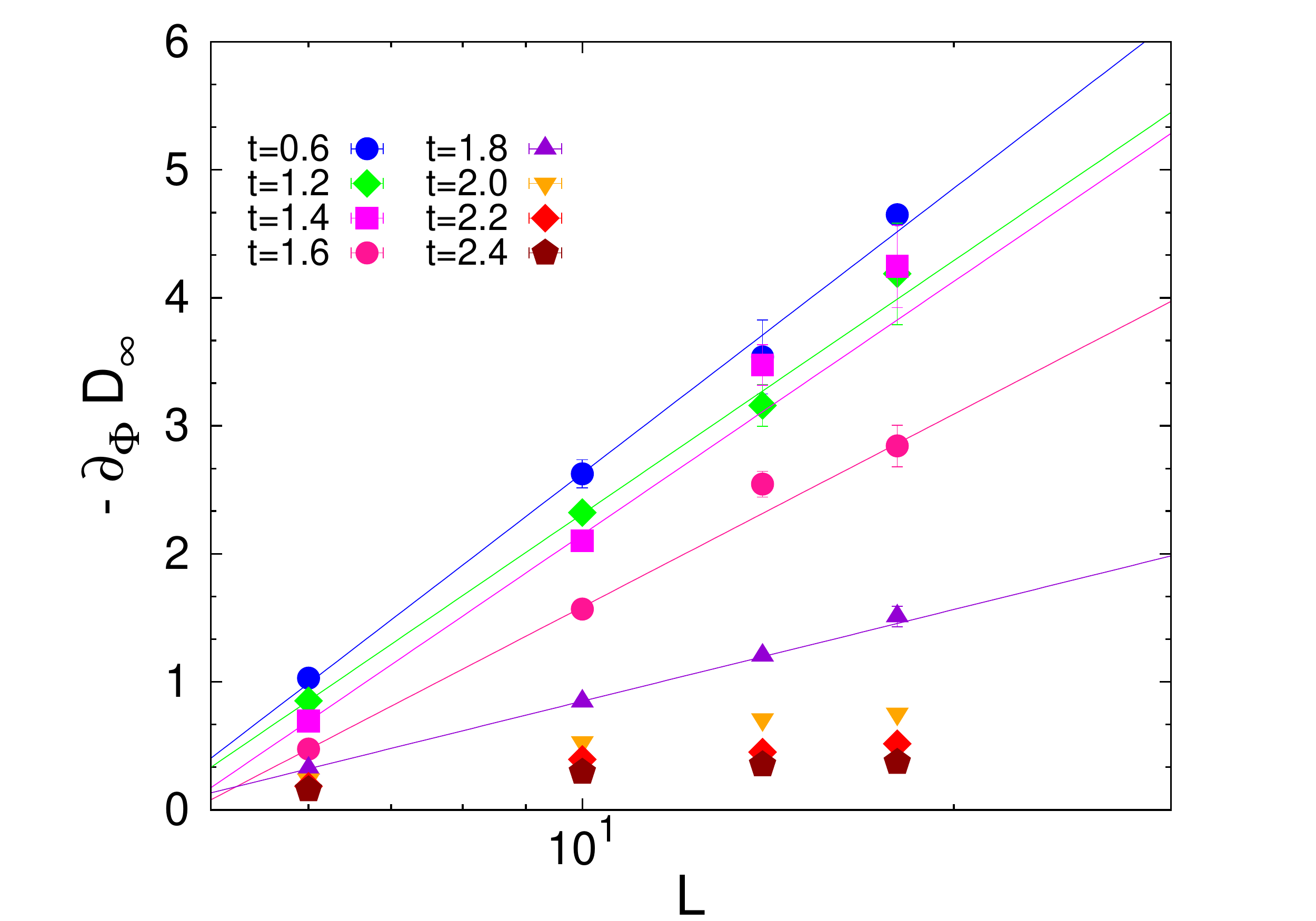}
 \caption{Scaling of $\partial_{\Phi} \langle D_{\infty} \rangle$ for the quantum rotor model on the cubic lattice with $U/(2Jn) = 4$ -- notice the logarithmic scale for $L$. Here $T_{\rm c}/(2Jn)=1.90(5)$. Solid lines are fits to the form $a\log L + b$.}
\label{f.Dnorm-CLscaling}
 \end{figure}

\subsubsection{Square and cubic lattice}

In the case of the square and cubic lattice, we focus on gauge-field thermometry based on the condensate peak asymmetry $D_{\infty}$ providing the highest accuracy as observed in the numerical simulations. 
As already mentioned in the main text, for the above lattices the Peierls phases introduced in Eq.~\eqref{e.QR} of the main text are gauge trivial, and they only induce a global shift of the momentum distribution: when applying a phase $\Phi$ one obtains the shift $n({\bm q}) \to n({\bm q} + \Phi \hat{x})$ (as shown in Fig.~\ref{f.squarecubic}(c)). Assuming that $\langle n({\bm q})/(N_+ + N_-)\rangle \approx \langle n({\bm q}) \rangle /\langle N_+ + N_-\rangle$, one obtains for the $\Phi$-derivative of the peak asymmetry $D_{\infty}$ at $\Phi=0$ the form
\begin{equation}
\partial_{\Phi} \langle D_{\infty} \rangle  \approx \frac{\sum_{q\in {\cal D}_+} \partial_{q_x} \langle n({\bm q}) \rangle - \sum_{q\in {\cal D}_-} \partial_{q_x} \langle n({\bm q}) \rangle}{\langle N_+ + N_-\rangle}  ~.
\label{e.dPhiDinfty}
\end{equation}
Here we have used the inversion symmetry of the momentum distribution around the $q_y$ axis for $\Phi=0$, leading to the identity
$\sum_{q\in {\cal D}_+} \partial_{q_x} \langle n({\bm q})\rangle = - \sum_{q\in {\cal D}_-} \partial_{q_x} \langle n({\bm q})\rangle$, which implies that $\partial_\Phi \langle N_+ - N_- \rangle=0$. It also implies that the scaling properties of  $\partial_{\Phi} \langle D_{\infty} \rangle$ are the same as those of each of the two terms composing the right-hand side of Eq.~\eqref{e.dPhiDinfty}.

\begin{figure}[h!]
 \includegraphics[width=8cm]{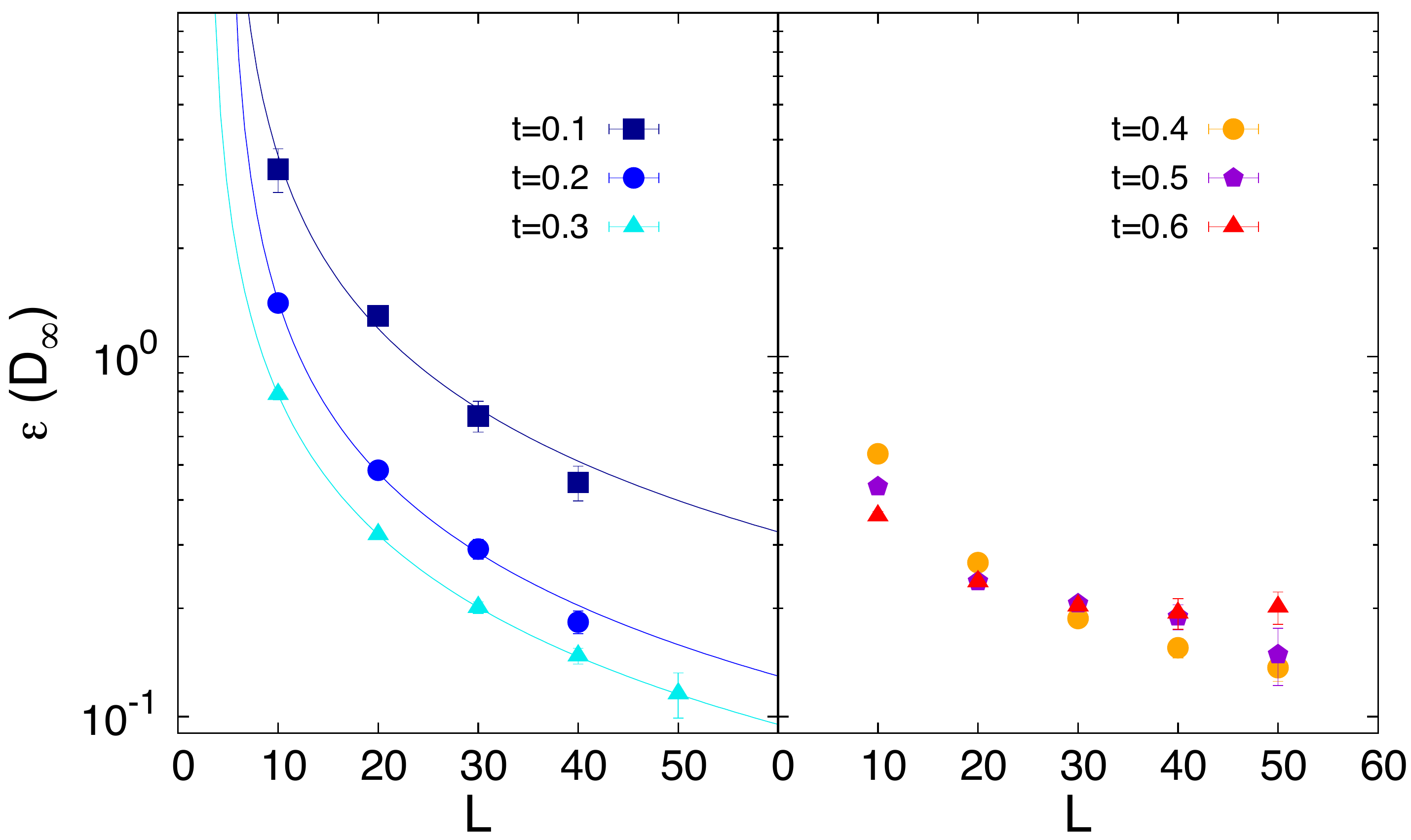}
 \caption{Scaling of the accuracy of the $D_{\infty}$ temperature estimator for the quantum rotor model on the square lattice with $U/(2Jn) = 5$.  Solid lines are fits to the form $a/(L+b)$.}
\label{f.eps-SLscaling}
 \end{figure}   

\begin{figure}[h!]
 \includegraphics[width=8cm]{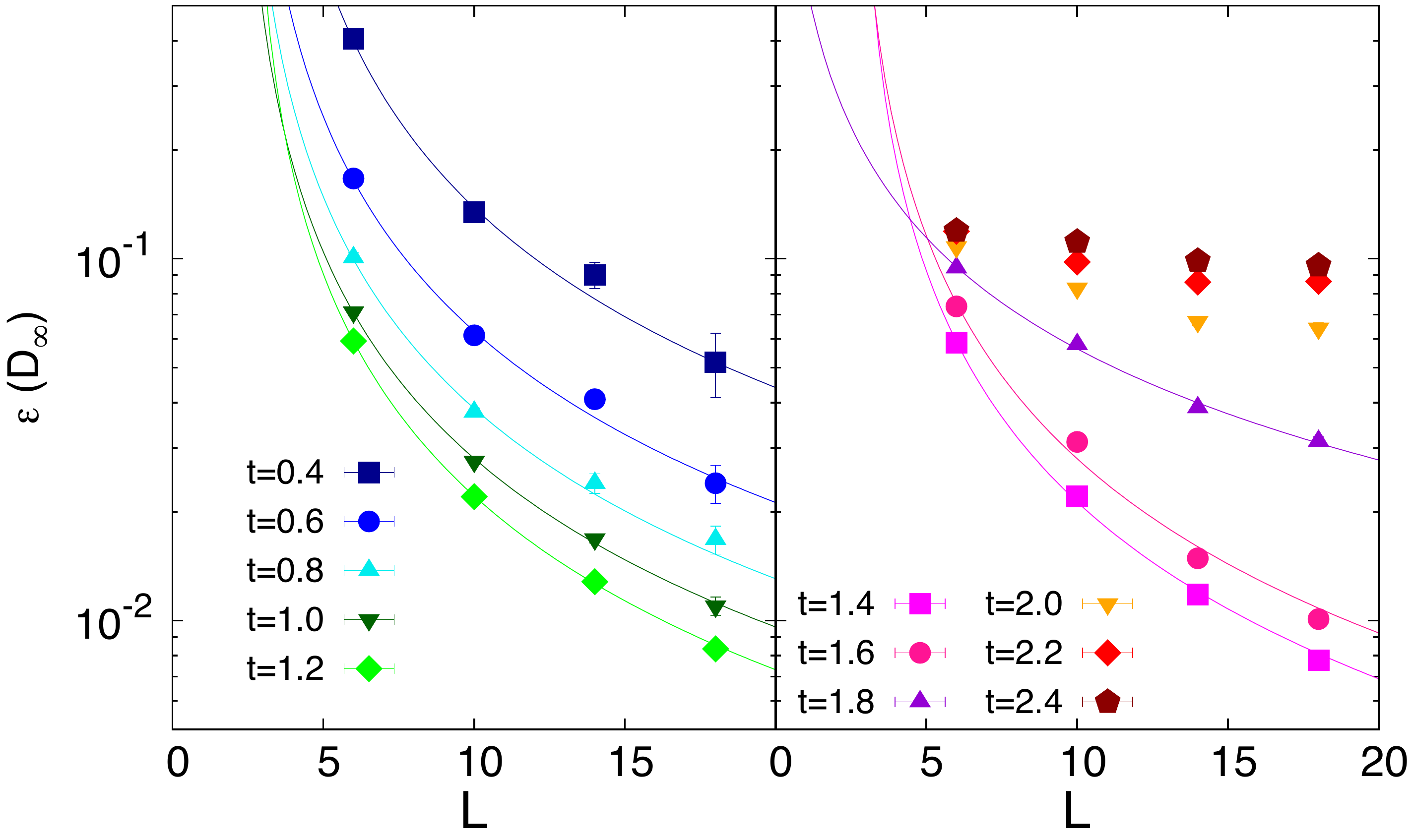}
 \caption{Scaling of the accuracy of the $D_{\infty}$ temperature estimator for the quantum rotor model on the cubic lattice with $U/(2Jn) = 4$. Solid lines are fits to the form $a [L(\log L + b)]^{-1}$.}
\label{f.eps-CLscaling}
 \end{figure}   

$D_{\infty}$ is sensitive to the $q-$dependence of $n({\bm q})$ around the condensation peak; in the condensate phase at low temperature, such a dependence is predicted by Bogolyubov theory to be $n({\bm q}) \sim q^{-2}$ \cite{PitaevskiiS2003}. We apply this prediction to the square and cubic lattice, for which the ${\bm q}-$vectors are discretized as $(2\pi/L)(n_1, n_2,...,n_d)$ (with $n_1, n_2, ..., n_d$ integers). Making use of Eq.~\eqref{e.dPhiDinfty} and of the fact that $\langle N_++N_-\rangle \sim L^d$ (given that it represents the integral of the momentum distribution over a finite fraction of momentum space), we obtain by numerical integration of $\partial_{q_x} n({\bm q})$ the dominant scaling behavior of $\partial_{\Phi} \langle D_\infty \rangle$ to be  
\begin{eqnarray}
\partial_{\Phi} \langle D_\infty \rangle   \sim & L ~~~~~&    (d=2) \nonumber\\
                                                            \sim & \log L ~~~~~&    (d=3)~. \nonumber
\end{eqnarray} 

Such a scaling behavior is indeed consistent with our numerical data for the quantum rotor model, as shown in Figs.~\ref{f.Dnorm-SLscaling} and \ref{f.Dnorm-CLscaling}, up to the Kosterlitz-Thouless (KT) transition in $d=2$ and up to the condensation transition in $d=3$.

 \begin{figure*}[ht!]
 \includegraphics[width=10cm]{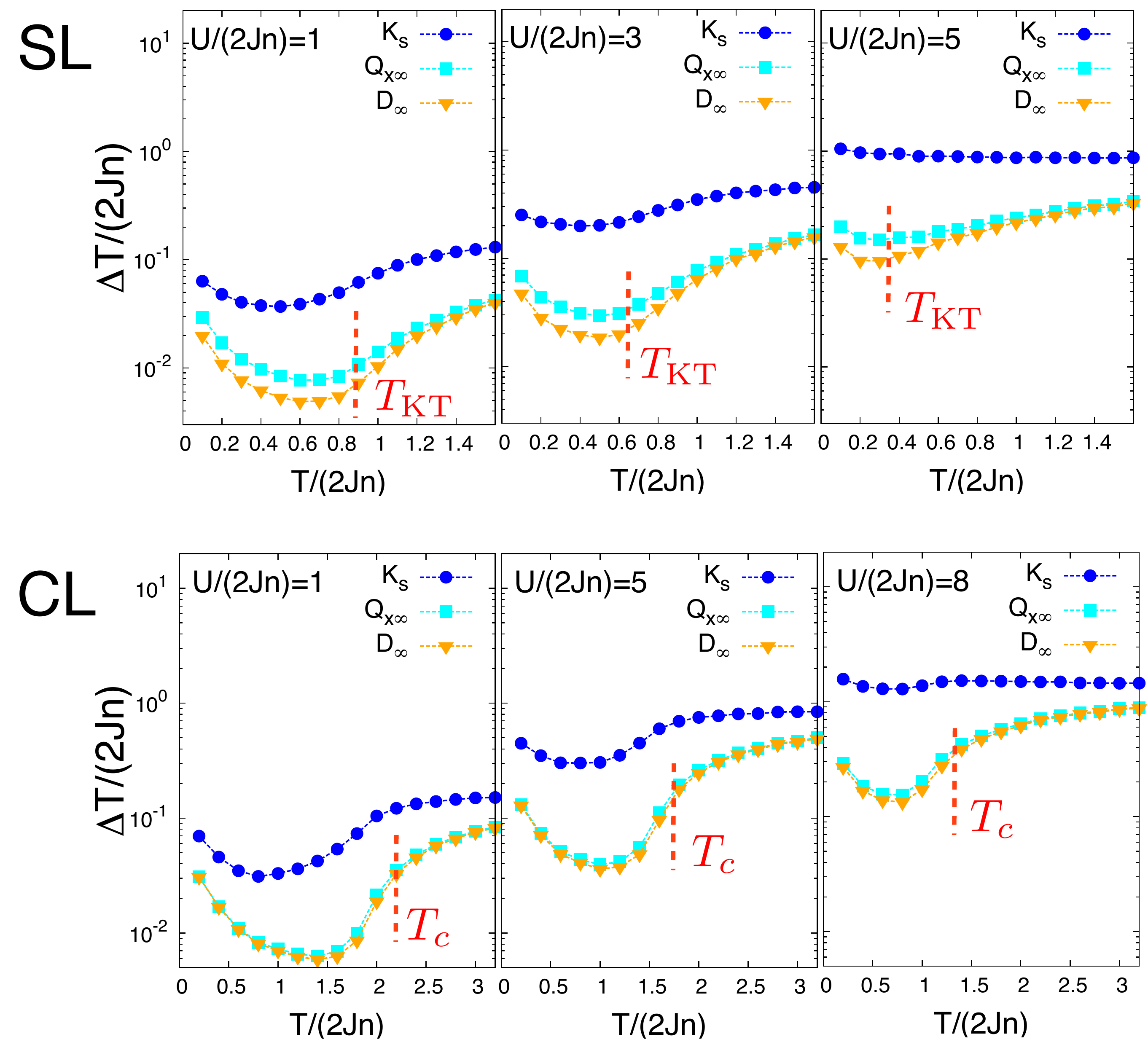}
 \caption{Absolute thermometry error $\Delta T = T_A - T$ for different estimators $T_A$ ($A = K_s, Q_{x\infty}, D_{\infty}$) and different interaction strengths in the quantum rotor model on the square lattice (SL - upper panels) and cubic lattice (CL - lower panels). The dashed lines indicate the Kosterlitz-Thouless critical temperature $T_{\rm KT}$ and the BEC critical temperature $T_c$. Square lattice: $T_{\rm KT}/(2Jn) = 0.85(4)$ ($U/(2Jn) = 1$), $T_{\rm KT}/(2Jn) = 0.64(2)$ ($U/(2Jn) = 3$),  $T_{\rm KT}/(2Jn) = 0.34(2)$ ($U/(2Jn) = 3$).
 Cubic lattice: $T_c/(2Jn) = 2.19(2)$ ($U/(2Jn)= 1$), $T_c/(2Jn) = 1.78(2)$ ($U/(2Jn)= 5$), $T_c/(2Jn) = 1.35(5)$ ($U/(2Jn)= 8$).}
\label{f.deltaT}
 \end{figure*}   

 \begin{figure}[h!]
 \includegraphics[width=8cm]{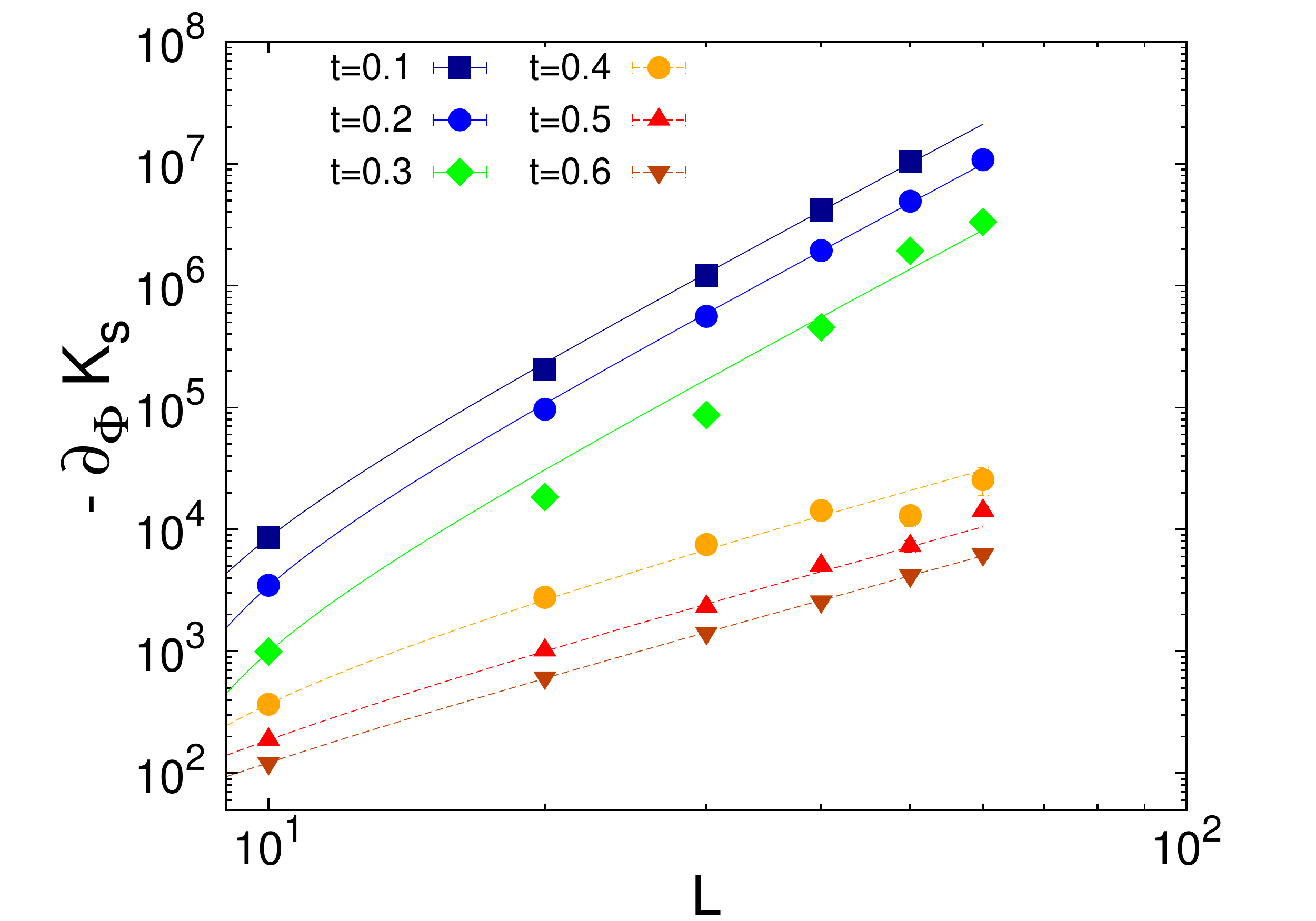}
 \caption{Scaling of $\partial_{\Phi} \langle K_s \rangle$ for the quantum rotor model on a triangular lattice with $\pi$-flux, $U/(2Jn) = 2$. For this model the chiral transition sits at $T_{\rm k}/(2Jn)=0.325(10)$. Solid lines are fits to the form $aL^4 + bL^2$; dashed lines are fits to the form 
$aL^2 + b$.}
\label{f.dKsdPhi-TLscaling}
 \end{figure}   

To reconstruct the full scaling of the relative error $\epsilon(D_{\infty})$ we would then need to extract the scaling of $\Delta{\rm cov}(D_{\infty},K_s)$. This appears to be rather challenging numerically. Indeed it turns out that, while the time-averaged and equal-time covariances grow with system size (as naturally expected), their difference appears not to scale with system size in $d=2$, or even to decrease with system size in $d=3$; hence the error bar on the difference grows much faster than the average, leading to very noisy results. 
On the other hand, the numerical estimate of the relative error $\epsilon(D_{\infty})$ obtained from $T_{D_\infty}$ is not affected by similar numerical problems, so that we rather reconstruct the scaling of the covariance difference from that of the relative error.  
Figs.~\ref{f.eps-SLscaling} and \ref{f.eps-CLscaling} show that, below the (quasi-)condensation transition, our numerical data are consistent with $\epsilon (D_\infty) \sim L^{-1}$ in $d=2$ -- implying a size-independent covariance difference --  and with $\epsilon (D_\infty) \sim (L \log L)^{-1}$ in $d=3$ -- implying a covariance difference scaling as $L^{-1}$. Remarkably, this very marked scaling of the thermometry error with system size is lost in the normal phase, showing that condensation is a necessary condition for gauge-field thermometry to achieve high accuracy in the presence of strong quantum fluctuations of the momentum distribution. Fig.~\ref{f.deltaT} further relates the accuracy of gauge-field thermometry with the onset of (quasi-)condensation: one observes that the \emph{absolute} thermometry error, $\Delta T = T(D_\infty)-T$ exhibits a marked decrease as a function of temperature in correspondence with the critical temperature for (quasi-) condensation.

\begin{figure}[h!]
 \includegraphics[width=8cm]{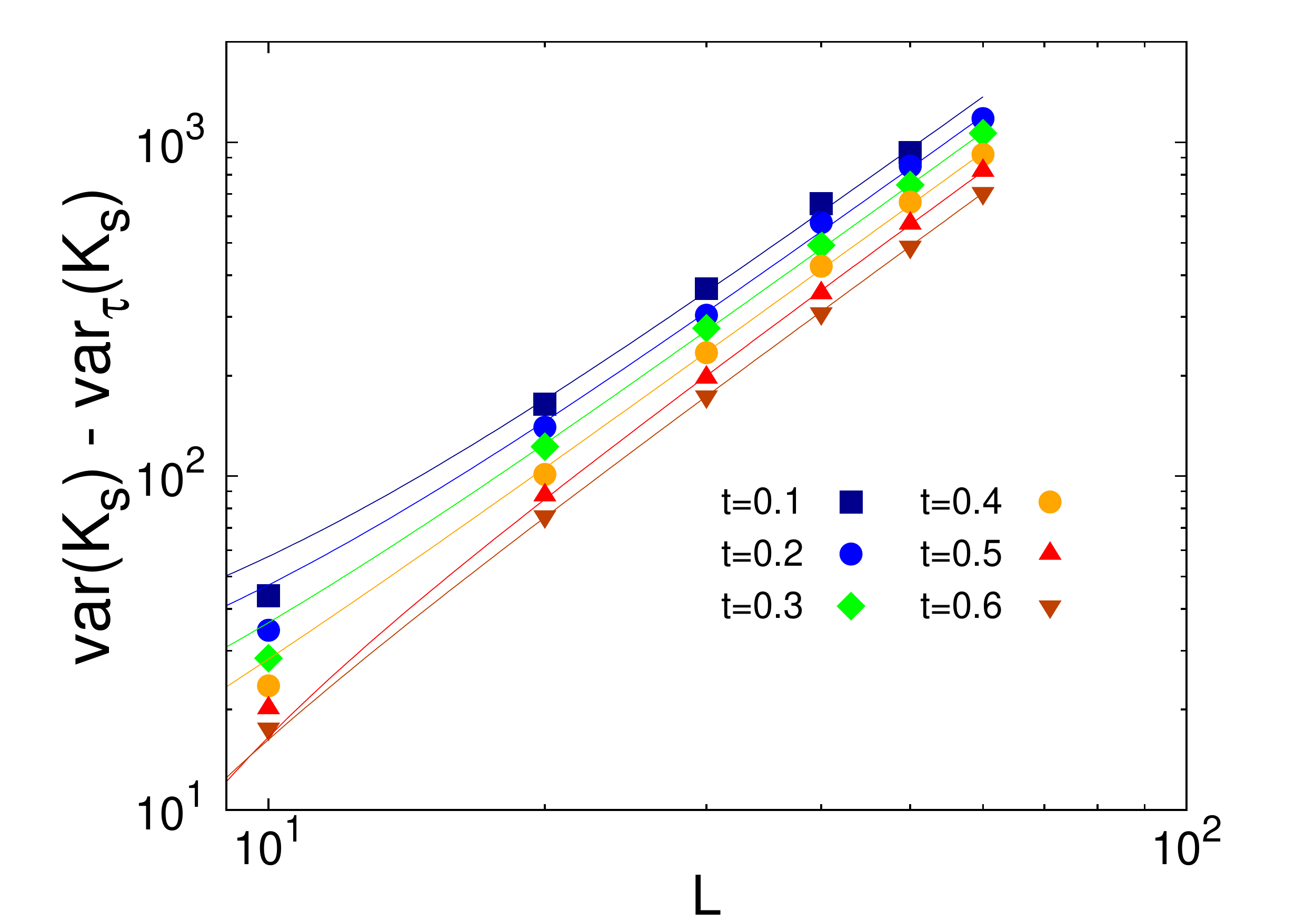}
 \caption{Scaling of the difference between equal-time and time-averaged $K_s$ variances for the quantum rotor model on a triangular lattice with $\pi$-flux, $U/(2Jn) = 2$. Solid lines are fits to the form $aL^2 + b$.}
\label{f.dvarKs-TLscaling}
 \end{figure} 
 \begin{figure}[ht!]
 \includegraphics[width=8cm]{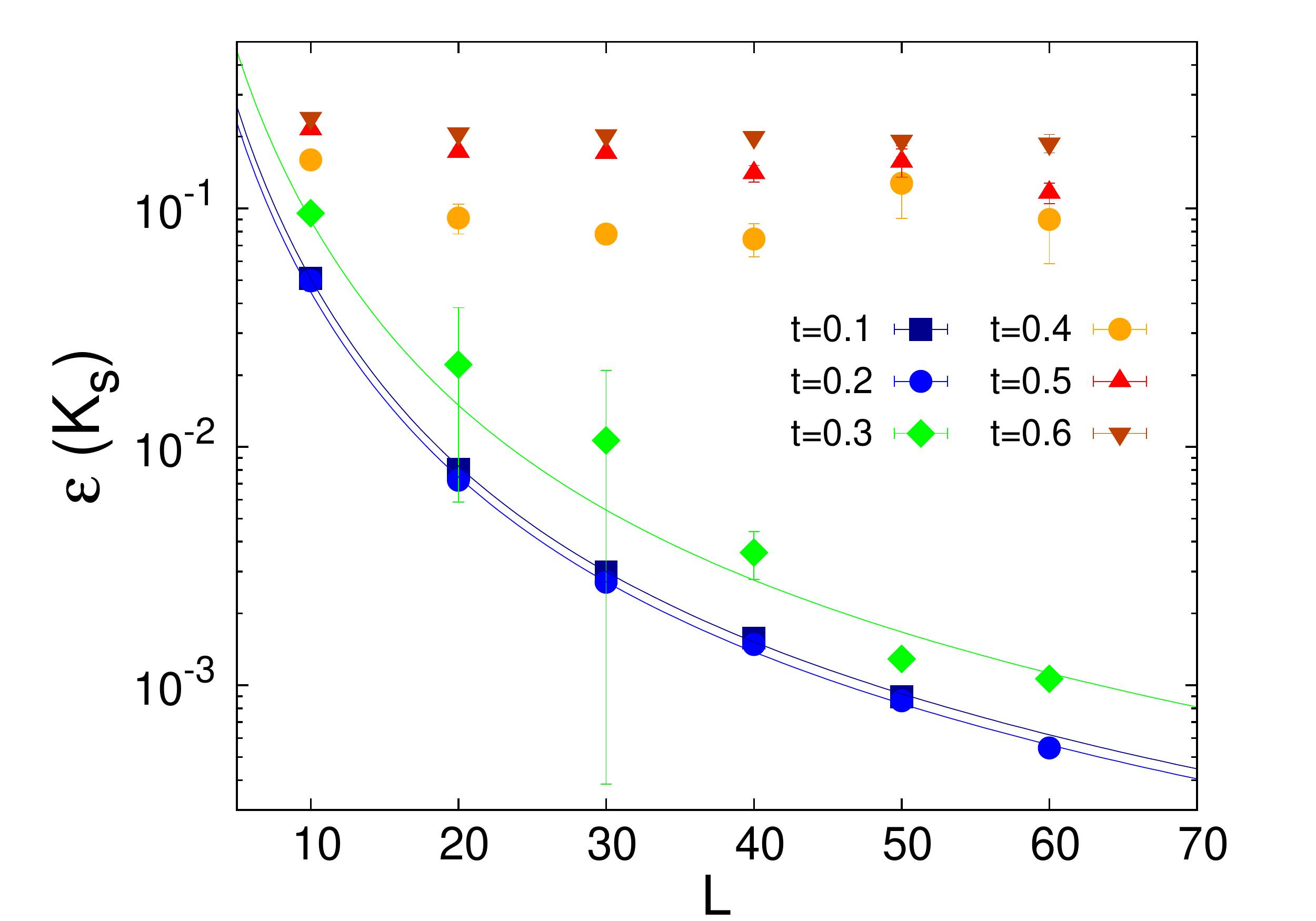}
 \caption{Scaling of the accuracy of the $K_s$ temperature estimator for the quantum rotor model on a triangular lattice with $\pi$-flux, $U/(2Jn) = 2$. Solid lines are fits to the form 
$a(b+ L^{-2})/(c+L^2)$.}
\label{f.eKs-TLscaling}
 \end{figure}   

\null\vspace*{2cm}
\subsubsection{Triangular lattice with $\pi-$flux}
In the case of the fully frustrated triangular lattice, as discussed in the main text the current operator $K_s$ acts as (extensive) Ising order parameter of the chiral transition. Therefore ${\rm var}(K_s) = \langle K_s^2\rangle - \langle K_s \rangle^2$ represents the equal-time order-parameter susceptibility, and as such it is expected to diverge with system size as $L^4$ in the ordered phase, as $L^{2+\gamma/\nu}= L^{15/4}$ at the critical point $T = T_k$, and as $L^2$ in the disordered phase. A similar scaling is expected for the time-averaged susceptibility, 
${\rm var}_{\tau}(K_s) = \frac{1}{\beta} \int_0^{\beta} d\tau \langle K_s(0)K_s(\tau)\rangle - \langle K_s \rangle^2$, which represents the actual response function to the variation of the applied flux:
\begin{equation}
\partial_\Phi \langle K_s \rangle = \frac{J}{T} ~{\rm var}_{\tau}(K_s)~.  
\end{equation}

All these expectations on the scaling of the variances of $K_s$ and of $\partial_\Phi \langle K_s \rangle$ are indeed confirmed by our numerical results, shown in Figs.~\ref{f.dKsdPhi-TLscaling} and \ref{f.dvarKs-TLscaling} for $U/(2Jn) = 2$.   
In particular the same scaling of 
${\rm var}(K_s)$ and ${\rm var}_{\tau}(K_s)$ applies also to their difference (Fig.~\ref{f.dvarKs-TLscaling}), entering in the definition of the relative error $\epsilon(K_s)$, Eq.~\eqref{e.epsilon}. The resulting scaling of $\epsilon(K_s)$ is then $\epsilon \sim L^{-2}$ for $T<T_k$, and $\epsilon$ independent of $L$ for $T>T_k$, as confirmed by Fig.~\ref{f.eKs-TLscaling}.


\begin{thebibliography}{99}
\bibitem{caution} This result, while valid for non-interacting fermions, ceases to be so for interacting bosons (on the same lattice), for which we observe that $\partial_\Phi \langle K_s \rangle=0$ for $\Phi \gtrsim \pi/2$.
\bibitem{footnotederiv} The existence of vanishing points for the above derivatives are expected by general symmetry arguments; nonetheless an experiment should be able to detect the poor accuracy of a temperature estimator from the strong decrease in the derivative of the associated observable.
\bibitem{footnoteD3d} In the case of a cubic lattice the ${\cal D}_{\pm}$ regions are actually cylinders in the three-dimensional momentum space, because we also account for integration along the line of sight ($z$ direction).
\bibitem{footnotepeak} Notice that the height of the condensate peak is \emph{insensitive} to the gauge field around $\Phi=0$, as the peak width is independent of $\Phi$, while the momentum occupation at $q=0$ attains a maximum for $\Phi=0$.
\end{thebibliography}
\end{document}